# A Physical Model for the Condensation and Decondensation of Eukaryotic Chromosomes.

Julien Mozziconacci*, Christophe Lavelle°, Maria Barbi#, Annick Lesne# and Jean-Marc Victor#.

*Cell Biology and Cell Biophysics Programme, European Molecular Biology Laboratory, Heidelberg, Germany
°CEA-DSV/DRR, Radiobiology and Oncology Group, 18, route du Panorama, 92265 Fontenay aux Roses Cedex, france
#Laboratoire de Physique Théorique de la Matière Condensée, CNRS UMR 7600,Université Pierre et Marie Curie, 4 Place Jussieu, 75252 Paris Cedex 05, France

## Abstract

During the eukaryotic cell cycle, chromatin undergoes several conformational changes, which are believed to play key roles in gene expression regulation during interphase, and in genome replication and division during mitosis. In this paper, we propose a scenario for chromatin structural reorganization during mitosis, which bridges all the different scales involved in chromatin architecture, from nucleosomes to chromatin loops. We build a model for chromatin, based on available data, taking into account both physical and topological constraints DNA has to deal with. Our results suggest that the mitotic chromosome condensation/decondensation process is induced by a structural change at the level of the nucleosome itself.

## Introduction

What happens to DNA during mitosis? Whereas biological and biochemical processes occurring during this crucial step of the cell cycle are rather well acknowledged, very few, if any, is known about the physical processes at work. The aim of this paper is to examine the physical constraints DNA has to deal with during the mitotic condensation and decondensation of eukaryotic chromosomes. Modern knowledge of the nucleus architecture assumes that DNA compaction is achieved by means of a hierarchical structure (Woodcock and Dimitrov). The first level of compaction is the helical wrapping of DNA around spools of proteins into nucleosomes (Luger). This first level of winding results in a "beads-on-a-string" assembly, which in turn folds into the so-called 30 nm chromatin fiber (Wolffe). This fiber is further organized into functional loops (Byrd, Labrador). This hierarchical structure has to perform extensive conformational rearrangements during the different steps of mitosis. Those rearrangement can be visualized, at the scale of the whole chromosomes be expressing GFP-tagged histones H2B in cells. After the duplication of the whole genetic material, mitotic chromosome condensation occurs during prophase. It has been shown that the formation of metaphasic chromosomes involves two distinct steps (Maeshima):
i.  Elongated chromatin fibers fold into a rod shaped structure.
ii. Both chromatids of each chromosome pair are resolved leading to the metaphasic well-known X-shaped chromosomes.

Sister chromatids are then segregated during anaphase to the daughter cell nuclei. During the re-establishment of the interphase nucleus architecture, each chromosome needs to decondense in order to allow transcription of some specific genomic regions. Recently, Manders *et al*. analyzed the dynamic behavior of chromatin during the transition from late anaphase to G1 in HeLa. They found that decondensation also occurs in two phases. First, a rapid decondensation by about a factor two of the entire chromosome occurs, followed by a slower phase in which part of the chromatin does not decondense any further, whereas the remaining chromatin decondenses further about two fold (Manders). At this point, it is important to notice that condensation involves the activity of ATP consuming enzymes and motors, whereas no ATP consumption is needed during decondensation.

In this paper we address the issue of how chromosomes achieve their compaction during prophase and decompaction after anaphase. The idea is to couple the different length scales involved in the conformational changes of the 30 nm fiber, integrating all the levels involved in the hierarchical organization of the genome, from the nucleosome up to the whole chromosome. The scenario we propose is based on physical considerations and focused on the topological and energetic constraints DNA has to deal with during the processes of compaction and decompaction.

## The physical requirements

A plausible model of the structural changes experienced by chromosomes during mitosis should be at least consistent with the following physical requirements:

i. In metaphasic chromosomes, nucleosomes have to be arranged in order to provide a DNA concentration as large as 0.20 g/mL (Daban). According to this concentration, the chromatin fiber must contain 10 nucleosomes per 10 nm: this implies the need for a space-filling fiber, with consequent steric hindrance constraints.

ii. Three relevant topological quantities, namely the twist Tw, the writhe Wr and the linking number Lk, which satisfy the identity Lk = Tw + Wr (Fuller), fully describe the DNA topology. The chromosome being organized into independent loops clamped at their ends, a safe and rapid unfolding of the metaphasic fiber has to preserve the DNA linking number Lk. This is a strong topological constraint that should be fulfilled.

iii. Beside topological constraints, energetics also has to be accounted for: the possible winding patterns that DNA can adopt in the fiber are strongly limited by the double helix stiffness. Indeed, bending and twisting persistence lengths (resp. 50 nm and 75 nm under physiological conditions) are significantly longer than the length of the linker (DNA linking two consecutive nucleosomes) which is of about 15-20 nm. As a consequence a curvature of 45° per linker or a twist rate of 35° per linker length (equivalent to 1 more bp per linker) already require an energy of $k_B T$ per linker.

iv. The precise winding pattern finally depends on electrostatics and specific (protein-protein and/or DNA-protein) interactions.

In order to evaluate the impact of these physical requirements on chromosome conformational changes we first need a model for the 30 nm chromatin fiber.

## A model for the fiber

Electron image and x-ray scattering data of isolated chromatin fibers from nuclei revealed that fibers are left-handed helices of stacked nucleosomes (Williams). Recently, both electron micrographs and digestion data obtained from regular nucleosome arrays reconstituted *in vitro* gave strong support to a two-start helix organization of the chromatin fiber (Dorigo).. Strikingly, the most recent model based on the X-ray structure of a tetranucleosome involves similar nucleosome and linker orientation (Schalch). Furthermore, a strong indication that linkers are cross-linking the fiber *in vivo* has been obtained using ionizing radiation on mitotic chromosomes (Rydberg). Therefore we here hypothesize that a zig-zag crossed-linker model is relevant for the fiber structure *in vivo*. We modeled such a fiber using the two-angle model of Woodcock *et al.* (Woodcock 93) as described in Barbi *et al.* (see Figure 1 e,g) and we previously showed (Barbi *et al.*) that this structure provides a maximal compaction of 6 nucleosomes per 10 nm. This compaction corresponds to the one found for isolated fibers *in vitro* (Williams) and may be assumed to be relevant *in vivo* during interphase. However, the linear density of the fiber in *metaphasic* chromosomes must be higher that the one of this latter structure. Namely, the fiber compaction must increase from 6 to 10 nucleosomes per 10 nm during the last compaction step of prophase (Daban). Different hypothetic fiber models have been proposed to account for the compaction observed in metaphasic chromosomes (Daban and Bermudez, Grigoriev). Neither of these models, however, agrees with the aforementioned experimental evidences.

**Our hypothesis:**
As regards this puzzle, we have recently proposed that an internal structural change of the nucleosome, which we named "gaping", could lead to the formation of a crossed-linker fiber compact enough to account for the DNA concentration in metaphasic chromosomes when starting from the interphasic crossed-linker fiber (Mozziconacci). The gaping structural change corresponds to detaching both H2A-H2B dimers from each other, leading to an opening of the nucleosome in the manner of a gaping oyster (see Figure 1 a,b). If this opening is accompanied by a twist of ~2 bp per linker, the external faces of neighboring nucleosomes come into close contact (Figure 1 c,d). The compaction of the fiber increases and reaches 10 nucleosomes per 10 nm (Mozziconacci) (Figure 1 e,f). We here hypothesize that this structural change is the key mechanism of chromatin condensation.

**Chromosome condensation as a topologically driven process**
The physical implications of this structural change have now to be analyzed, in order to check its feasibility. Let's start by considering the topological effect of nucleosome gaping on a fiber.
The comparison between the fiber structures displayed in figures g and h shows that the gaping of all nucleosomes within the fiber results in a twist of the fiber around its own axis. In order to describe this twist properly, one has to refer to the precise calculation of the topological properties of DNA in the fiber. We have calculated the linking number of DNA in all the allowed conformations of the fiber (Barbi), as obtained using the canonic two-angle model of Woodcock *et al.* (Woodcock). On the basis of this calculation, we have shown that it is possible to define a twist and a writhe for any 30 nm fiber structure:

i. The fiber writhe, $Wr^F$, results to be simply the writhe of its local axis;
ii. Its twist, $Tw^F$, can be identified with the rotation angle between two consecutive nucleosomes along the fiber axis, in the same way as the twist of DNA is defined by the rotation angle of one base pair relative to the previous one ;
iii. Moreover, these two quantities share the same properties as the twist and writhe of the DNA double helix, but *at the level* of the chromatin fiber: their sum $Wr^F + Tw^F$ is equal to the linking number of the fiber, $Lk^F$, which is itself equal to the linking number of the DNA up to a constant[1](Barbi).

This formal characterization of the fiber topology allows us to calculate the change in the fiber twist induced by the gaping of all the nucleosomes. We found it equal to 7 turns for 1000 nucleosomes. This means that during the compaction of a fiber loop containing ~200 kbp, i.e. ~1000 nucleosomes, the nucleosome at one end should rotate around the fiber axis by ~7 turns with respect to the one at the other end. However both ends of the loop being fixed at clamped boundaries, such a rotation is forbidden. Hence, in order to preserve its global linking number $Lk^F$, the compact fiber has to be writhed at an inverse rate of ~ -7 turns per loop. Depending on the geometry of the loops in the network of chromatin fibers and on the bending energy of the fiber, this compensatory writhe can either result in a toroidal or a plectonemic supercoiling (see figure g, h, i, j). We have evaluated the fiber writhe for these two coiling models. For a plectoneme, depending on the length of the end loop, it ranges from -7 to -5. For a toroidal solenoid, it ranges from -7 to - 6 depending on its pitch and its radius. Remarkably, these changes in the writhe perfectly match the twist induced by the gaping of all nucleosomes within the loop. We therefore propose that nucleosome gaping may not only compact the fiber, but may also be the driving mechanism for supercoiling the fiber loop in a condensed higher order structure.

---
[1] This constant is the DNA twist in a straight and relaxed state.

**Chromosome condensation is an active process**
Before looking at the decondensation process, we qualitatively address the energetic imbalance of the compaction process. Gaping of all nucleosomes within a chromatin loop requires energy (Mozziconacci). The un-sticking of both H2A-H2B dimers from each other requires 30 kcal/mol. Nevertheless, the repulsion between the two apposing gyres of the nucleosomal DNA reduces the net energy cost of about ~15 kcal/mol in physiological conditions (Kulic). Furthermore, linkers are twisted at ~2 bp per linker in order to stack neighboring nucleosomes in the fiber, this resulting in an additional energy (~4 $k_B T$, or 2.4 kcal/mol) to the gaped state. On the other hand the stacking of the external faces of neighboring nucleosomes provides an energy gain of ~20 kcal/mol from which the electrostatic contribution of ~15 kcal/mol must be again cut off. Figure 2 displays the resulting energy profile. This stacking involves the formation of ionic bridges between neighbouring nucleosome faces thanks to divalent cations, as shown in crystallographic structures (Davey). Very interestingly, a major influx of divalent cations ($Ca^{2+}$, $Mg^{2+}$) has been evidenced in the first stages of mitosis (Strick). In this scenario, the compaction process corresponds to crossing an energy barrier to go from a stable (interphasic) to a metastable (metaphasic and gaped) state. An active, ATP consuming, process should supply the energy required for crossing the barrier. One can hypothesize that condensins, which are known to induce the supercoiling of a DNA plasmid in vitro (Kimura), are part of the molecular motors at work during this process (Legagneux, Gassmann). We note that in the earlier stages of mitosis, other molecular motors, namely remodeling enzymes, have been shown to be involved in the removal of all transcription factors and the redistribution of nucleosomes into regular arrays (Komura).

**Chromosome decondensation and formation of the interphase nucleus**
In our model, nucleosome gaping should be considered therefore as an energy consuming process that allows the fiber to simultaneously condense and coil during prophase. The resulting metaphasic chromosome has been found to be metastable. We then argue that the stored energy should be used, at the end of mitosis, to drive the fiber back to its more stable interphasic configuration. This transition is, indeed, energetically favorable, and can take advantage of the high cooperativeness of the process: if one nucleosome is un-gaped, then its two neighboring nucleosomes will not be perfectly stacked anymore, and then will tend to switch back into their closed form too. The un-gaping of the fiber can thus spread all over the chromosome or at least until a loop boundary is reached.
Of course, nucleosome un-gaping still requires some energy, because an energy barrier has to be crossed (see Figure 2). Very interestingly, the barrier height can be further tuned by the ionic strength: it influences indeed the DNA-DNA repulsion quite strongly, while having little or no effect on the protein-protein interaction, and then on the energy levels of the stable and metastable states (Sun). Therefore, decondensation can be favored by a change in the biophysical and biochemical environmental parameters.
In our model, the storage of mechanical energy in the gaped fiber can result, once the gaping constraint released, in mechanical forces acting toward a rapid fiber elongation (by a factor ~2) and loop uncoiling. This is in keeping with the observation of a first rapid phase in the decondensation process as mentioned in the introduction (Manders).
At the end of this first decondensation, no substantial chromatin region remains as dense as in late anaphase (Weidemann). Therefore, in our hypothesis, nucleosomes are likely to be all un-gaped at this stage. Then, as the chromatin fiber is not locked anymore by nucleosome stacking interactions, a further decondensation can occur at specific regions.

A possible mechanism for this decondensation, characterized by a strong elongation of the fiber and the conservation of the DNA linking number, has already been described (Barbi). During this elongation, the linear density of the fiber can decrease from 6 to 2 nucleosomes per 10 nm. The totally decondensed state is likely to be ready for transcription. It seems reasonable that each chromosomal loop is then either totally decondensed or not at all. This specific decondensation, regulated by biochemical modifications, finally leads to the formation of the interphase nucleus, with still dense regions corresponding to heterochromatin and less dense regions corresponding to euchromatin (Cremer, O'Brien, Verschure).

## **Conclusion**

In summary, we propose that a conformational change in the nucleosome structure (Mozziconacci) may be the key mechanism driving the process of condensation and decondensation of the 30 nm chromatin fiber during mitosis and analyze the physical and biological implication of this assumption. Our model involves two condensation/decondensation steps, in keeping with most recent findings.

One important consequence of our hypothesis for the kinematics of chromatin during mitosis is that in eukaryotes, condensation/decondensation of chromosomes may be achieved at constant DNA linking number by changing the internal parameters defining the fiber geometry. This consideration implies that the late condensation and the early decondensation of chromosomes can occur without the intervention of topoisomerases' *catalytic* activity (experimental evidence tends to support this result, see e.g. Lavoie). Concerning the decondensation process, it has been shown in vivo that chromatin decondensation could take place even when those enzymes were inhibited (Wright and Schatten). Furthermore, a classical experiment on SV40 chromatin clearly demonstrates that the DNA linking number is not affected by the decondensation process in vitro (Keller). Besides, one can imagine that decondensation mechanisms not involving these enzymes could be favorable for the cell, in terms of efficiency and safety. The question therefore arises whether chromatin may have evolved to limit topoisomerase activity during the unwinding process. The scenario proposed in this paper, based on energetic and topological considerations, supports this conjecture. More generally, we suggest that the analysis of physical constraints may be useful to understanding biological processes.

* corresponding author: mozzicon@embl.de

# FIGURE CAPTIONS

Fig. 1. Coiling of the 30 nm chromatin fiber induced by nucleosome gaping.
We present here models of nucleosome and fiber before gaping (on the left) and after gaping (on the right). Each color box refers to a particular length scale.
**a and b**: Molecular model illustrating the gaping process, i.e. the hinge opening of the nucleosome around an axis crossing the dyad axis and bridging the cysteines 110 of both histones H3.
**c and d**: Three consecutive nucleosomes in a fiber. The gaping process induces the perfect stacking of two neigboring nucleosomes.
**e to h**: side views (**e**, **f**) and top view (**g**, **h**) of the chromatin fiber. The DNA of the top nucleosome has been highlighted in cyan. A yellow (resp. orange) arrow represents its position relative to the bottom one before (resp. after) gaping. The change in the twist of the fiber is equal to the angle pointed out by the blue arrow.
**i, j, k, l**: Models of chromatin loop before and after gaping. Depending on the geometry of the chromosome network, loops can either coil into solenoids (j) or plectonemes (l).

Fig. 2: Energy profile associated with the gaping transition.
Sketch of the energy profile describing the transition between the un-gaped (left) and the gaped (right) nucleosome in the fiber. Energies are indicated in kcal/mol. The overall energy difference between the two states, of about 10 kcal/mol, results from the balance between inter- and intra-nucleosomal sticking energies (blue arrows). Electrostatic repulsion between DNA gyres (red arrows) does not affect the energy of the two states but decreases the height of the activation barrier between them.

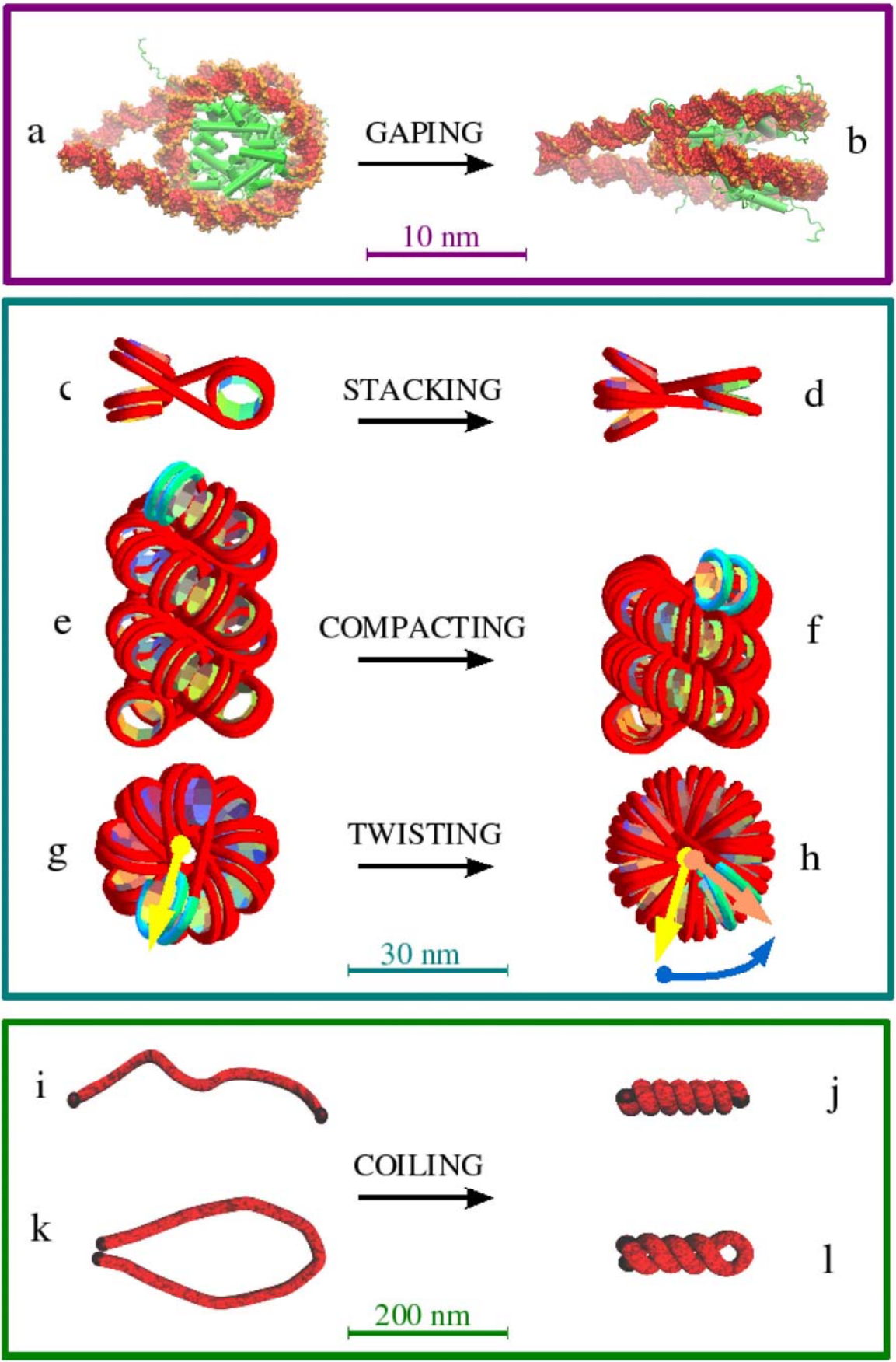

Fig. 1

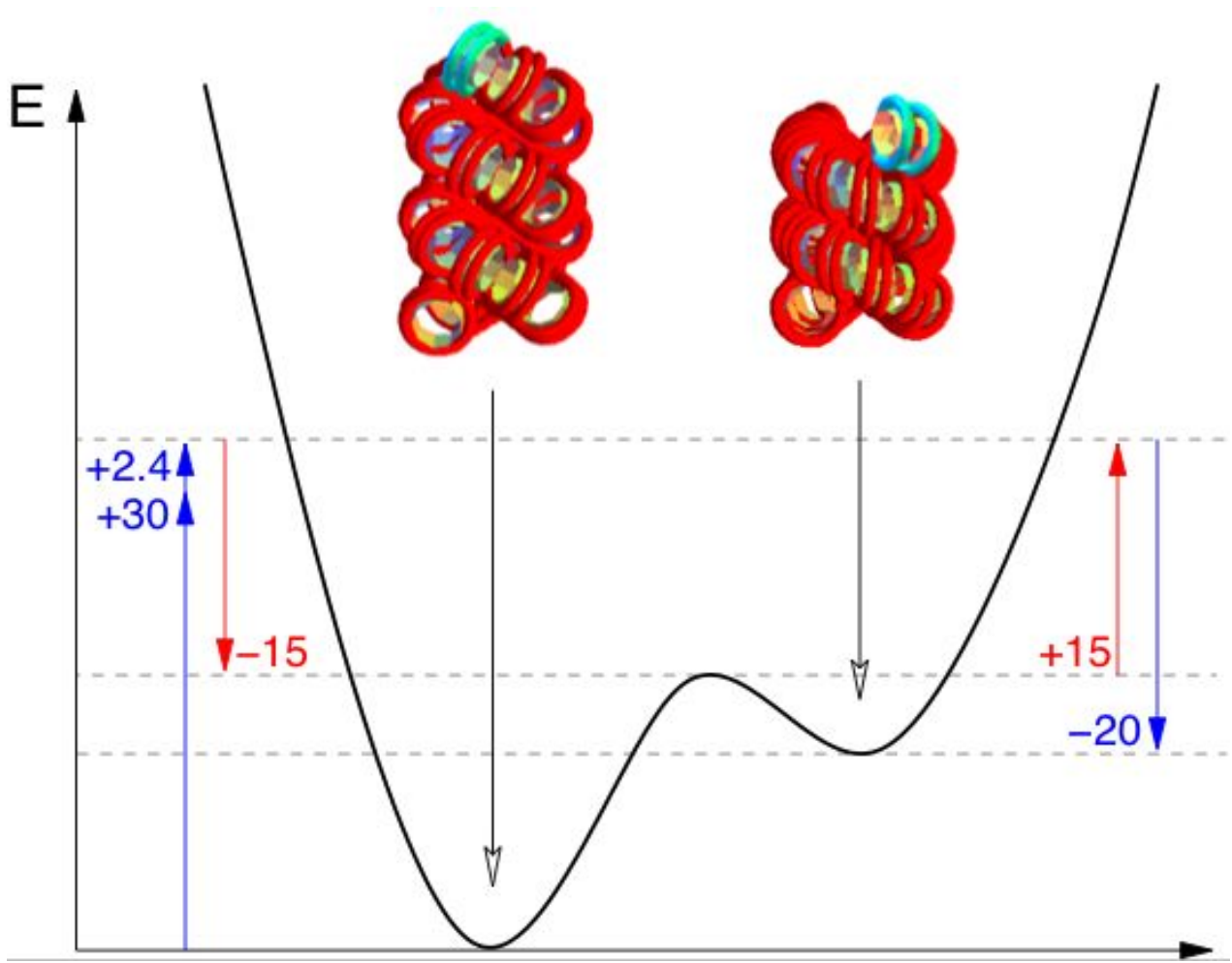

Fig. 2